\documentclass{emulateapj}

\usepackage{natbib}

\usepackage{graphicx}
\usepackage{latexsym}
\usepackage{amsfonts,amsmath,amssymb}
\usepackage{url}
\usepackage[utf8]{inputenc}
\usepackage{fancyref}
\usepackage{hyperref}
\hypersetup{colorlinks,breaklinks,
            urlcolor=[rgb]{1,0,0},  
            linkcolor=[rgb]{0,0,0},       
            citecolor=[rgb]{0,0.4,1.0}}   

\newcommand{\varv}{v}

\newcommand{\kms}{\, {\rm km}\, {\rm s}^{-1}}

\newcommand{\llso}{\log\, L/ L_\odot \,}
\newcommand{\simle}{\mathrel{\hbox{\rlap{\hbox{\lower4pt\hbox{$\sim$}}}\hbox{$<$}}}}
\newcommand{\simgr}{\mathrel{\hbox{\rlap{\hbox{\lower4pt\hbox{$\sim$}}}\hbox{$>$}}}}

\newcommand{\code}{}

\newcommand{\ADIPLS}{\code{ADIPLS }}

\newcommand{\KIC}{\rm{KIC}8366239}
\newcommand{\KID}{\rm{KIC}5006817}

\newcommand{\unitspace}{\ensuremath{\,}}
\newcommand{\usp}{\unitspace}

\newcommand{\unitstyle}[1]{\ensuremath{\mathrm{#1}}}
\newcommand{\power}[2]{\ensuremath{{#1}^{#2}}}

\newcommand{\centi}{\unitstyle{c}}

\newcommand{\meter}{\unitstyle{m}}

\newcommand{\cm}{\centi\meter}
\newcommand{\gram}{\unitstyle{g}}

\newcommand{\GramSc}{\gram\usp\power{\cm}{2}}

\newcommand{\Msun}{\ensuremath{\unitstyle{M}_\odot}}
\newcommand{\Lsun}{\ensuremath{\unitstyle{L}_{\odot}}}
\newcommand{\Rsun}{\ensuremath{\unitstyle{R}_{\odot}}}

\newcommand{\cms}{\,  {\rm cm}^{2}\, {\rm s}^{-1}}

\newcommand{\Teff}{\ensuremath{T_{\!\mathrm{eff}}}}	
\newcommand{\teff}{\Teff}

\newcommand{\alphaMLT}{\ensuremath{\alpha_{\mathrm{MLT}}}}	
\newcommand{\alphasc}{\ensuremath{\alpha_{\mathrm{sc}}}} 

\newcommand{\mso}{\Msun}
\newcommand{\rso}{\Rsun}
\newcommand{\lso}{\Lsun}

\shorttitle{Angular momentum transport within evolved low-mass stars}
\shortauthors{Cantiello et al. 2014}

\begin{document}

\title{Angular momentum transport within evolved low-mass stars}

\author{
Matteo Cantiello\altaffilmark{1},
Christopher Mankovich\altaffilmark{2,3},
Lars Bildsten\altaffilmark{1,3},\\
J{\o}rgen Christensen-Dalsgaard\altaffilmark{4}
and
Bill Paxton\altaffilmark{1}
}

\altaffiltext{1}{Kavli Institute for Theoretical Physics, University of California, Santa Barbara, CA 93106, USA}
\altaffiltext{2}{Department of Astronomy and Astrophysics, University of California, Santa Cruz, CA 95064, USA}
\altaffiltext{3}{Department of Physics, University of California, Santa Barbara, CA 93106, USA}
\altaffiltext{4}{Stellar Astrophysics Centre, Department of Physics and Astronomy, Aarhus University, Ny Munkegade 120, DK-8000 Aarhus C, Denmark}
\email{matteo@kitp.ucsb.edu}

\begin{abstract}
Asteroseismology of $1.0-2.0M_\odot$ red giants by the \emph{Kepler}
satellite has enabled the first definitive measurements of interior
rotation in both first ascent red giant branch (RGB) stars and those
on the Helium burning clump. The inferred rotation rates are $10-30$
days for the $\approx 0.2M_\odot$ He degenerate cores on the RGB and
$30-100$ days for the He burning core in a clump star. Using the MESA
code we calculate state-of-the-art stellar evolution models of low
mass rotating stars from the zero-age main sequence
to the cooling white dwarf (WD) stage. We include transport of angular
momentum due to rotationally induced instabilities and circulations, as well as  magnetic fields in radiative
zones (generated by the Tayler-Spruit dynamo). We find that all models
fail to predict core rotation as slow as observed on the RGB and
during core He burning, implying that an unmodeled angular momentum
transport process must be operating on the early RGB of low mass
stars. Later evolution of the star from the He burning clump to the
cooling WD phase appears to be at nearly constant core angular
momentum. We also incorporate the adiabatic pulsation code, ADIPLS, to
explicitly highlight this shortfall when applied to a  specific
\emph{Kepler} asteroseismic target, KIC8366239. 
\end{abstract}


\bibliographystyle{apj}

\section{Introduction}
Stellar rotation and the resulting internal rotational profile, together with the mechanisms that contribute to angular momentum transport, remain poorly probed.
Different classes of transport mechanisms have been proposed, in particular hydrodynamical instabilities and circulations induced by rotation \citep[see][for a review]{Maeder:2000}, magnetic torques \citep{Gough:1998,Spruit:1999,Spruit:2002,Spada:2010} and internal gravity waves \citep[see e.g.,][]{Charbonnel:2005}. In the absence of strong mass loss, the bulk of the redistribution of angular momentum is expected to occur when shearing is generated during evolutionary episodes of expansion or contraction. 

Most stars ignite hydrogen in a shell at the end of the main sequence. Above this shell the star begins to expand, while the core contracts. In the absence of a strong coupling between the core and the envelope, conservation of angular momentum requires that the core spins up while the envelope spins down. This implies very rapidly rotating stellar cores and a shear layer between core and envelope. Evolutionary calculations that include angular momentum transport from rotational instabilities and circulations also predict rapidly rotating stellar cores at the end of stellar evolution \citep[e.g.,][]{Heger:2000,larends_Yoon_Heger_Herwig_2008,Eggenberger:2012,Marques:2013}. This is at odds with the observed rotation rate of white dwarfs (WD) and neutron stars (NS). This  can be somewhat remedied by including angular momentum transport due to magnetic torques in radiative regions \citep[Tayler-Spruit dynamo (TS)][]{Spruit:2002,Heger:2005,larends_Yoon_Heger_Herwig_2008}. 

In the era of space asteroseismology, the \textit{Kepler} satellite enabled the measurement of the core rotation in many red giant branch (RGB) stars using  the splitting of mixed modes \citep{Beck:2012,Deheuvels:2012,Mosser:2012,Deheuvels:2014}. Mixed modes are oscillations that have an acoustic component (p-mode) in the envelope and are g-modes (restoring force is buoyancy) in the  stellar core \citep[see e.g.,][]{Beck:2011}.
\citet{Mosser:2012a,Mosser:2012} showed that it is the rotation rate
 of the material below the active hydrogen burning shell that is most directly inferred from 
the  splitting of mixed modes \citep[see also][]{Marques:2013}. This  measurement 
of the interior rotational state of an evolved star provides a new test for  
theoretical ideas of angular momentum transport.
Similar to the case of compact remnants, models solely including angular momentum transport due to rotational mixing and circulations predict rotation rates 2 to 3 orders of magnitude higher than observed \citep{Eggenberger:2012,Marques:2013,Ceillier:2013}. While the adopted treatment of angular momentum transport is a crude approximation, \citet{Marques:2013} have shown that even the most extreme of the physically motivated hydrodynamic mechanisms included in their code cannot yield the observed slow rotation.

Our goal is to assess whether models including transport due to TS magnetic fields agree more closely with the  observations.
While the physics of the Tayler instability is secure, the existence of the Tayler-Spruit dynamo loop is debated on both analytical and numerical grounds \citep{Braithwaite:2006,Zahn:2007}. However observations of the spin rates of compact objects (WDs and NSs) are in much better agreement with models including this angular momentum transport mechanism \citep{Heger:2005,larends_Yoon_Heger_Herwig_2008}, which has also been discussed in the context of the rigid rotation of the solar core \citep{Eggenberger:2005}, but see also \citet{Denissenkov:2010}.
In Sec.~\ref{stev} we present the stellar evolution calculations and the details of the implemented physics. 
Results for the evolution of core rotation during the early RGB are shown in  Sec.~\ref{earlyrgb} for different angular momentum transport mechanisms. Results are compared to \textit{Kepler} asteroseismic observations of mixed modes in RGB stars. In Sec.~\ref{splittings} we show how the rotational splittings of mixed modes are calculated from the stellar evolution models using \ADIPLS  \citep{JCD:2008}. 
Sec.~\ref{beyond} presents the angular momentum evolution of our models beyond the early RGB. Predictions for the core rotation rates past the luminosity bump (Sec.~\ref{bump}), during core He-burning  (Sec.~\ref{clump})  and in the WD stage  (Sec.~\ref{wd}) are shown and compared to the asteroseismically derived values. 
In Sec.~\ref{conclusions} we draw our conclusions and discuss possible future work.

\section{Stellar evolution calculations}\label{stev}
We use the Modules for Experiments in Stellar Evolution (MESA, release 5118) code to evolve low-mass stars from the pre-main-sequence to the cooling WD sequence \citep{Paxton:2011,Paxton:2013}. This code includes the effects of the centrifugal force on the stellar structure, chemical mixing and transport of angular momentum due to rotationally induced hydrodynamic instabilities \citep{Heger:2000}. The mixing of angular momentum due to dynamo-generated magnetic fields in radiative zones is also included \citep{Spruit:2002,Petrovic:2005,Heger:2005}. See \citet{Paxton:2013} for the details of the implementation of rotation and magnetic fields in MESA.

We chose an initial metallicity of $Z=0.02$ with a mixture taken from \citet{Asplund:2005}. We adopt the OPAL opacity tables \citep{Iglesias:1996}  accounting for the carbon- and oxygen- enhanced opacities during helium burning \citep[Type 2 OPAL,][]{Iglesias:1993}.
Solid body rotation is set at the zero-age main sequence (ZAMS). 

\begin{figure}[htp!]
\begin{center}
\includegraphics[width=1\columnwidth]{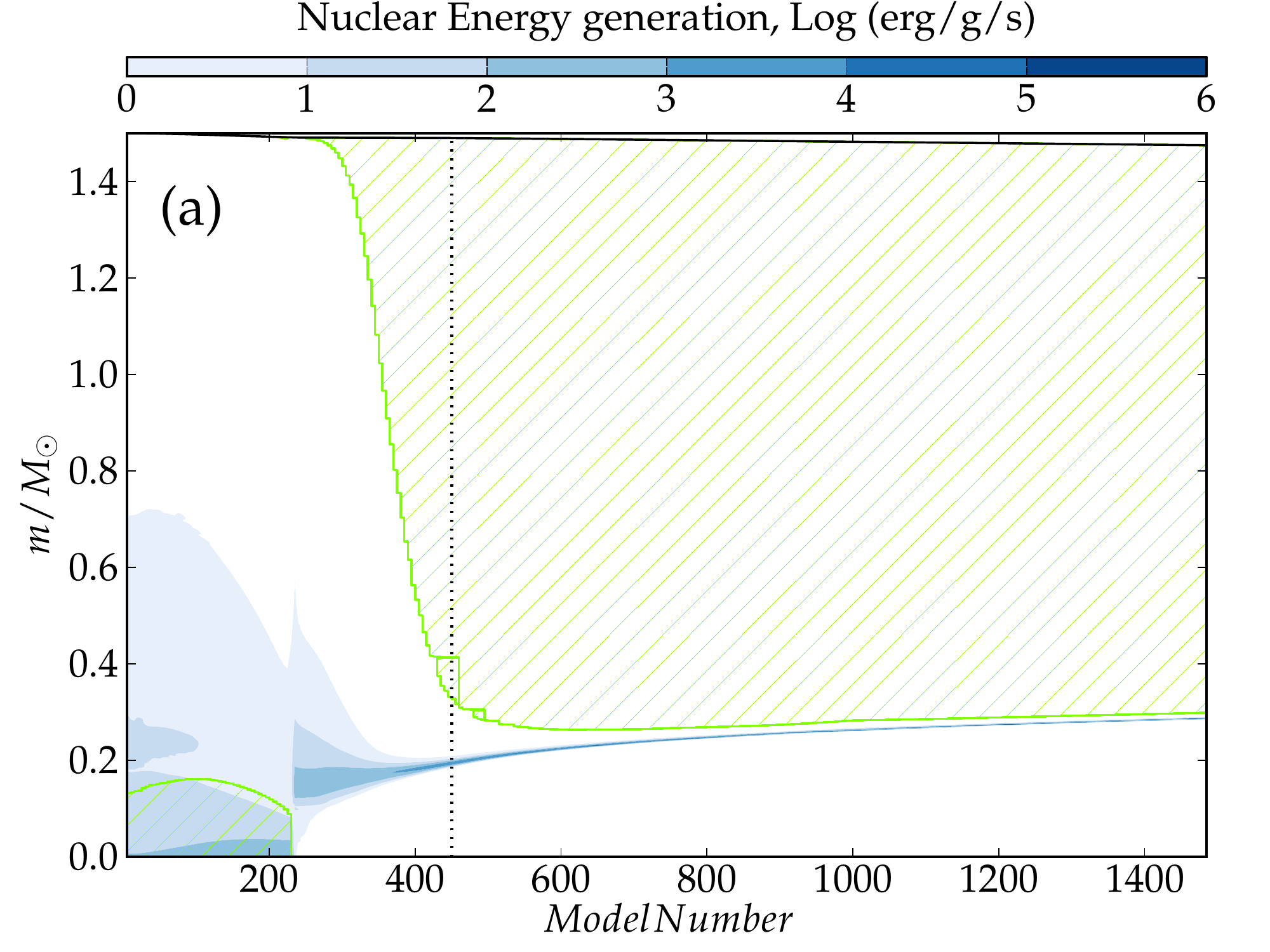}
\includegraphics[width=1\columnwidth]{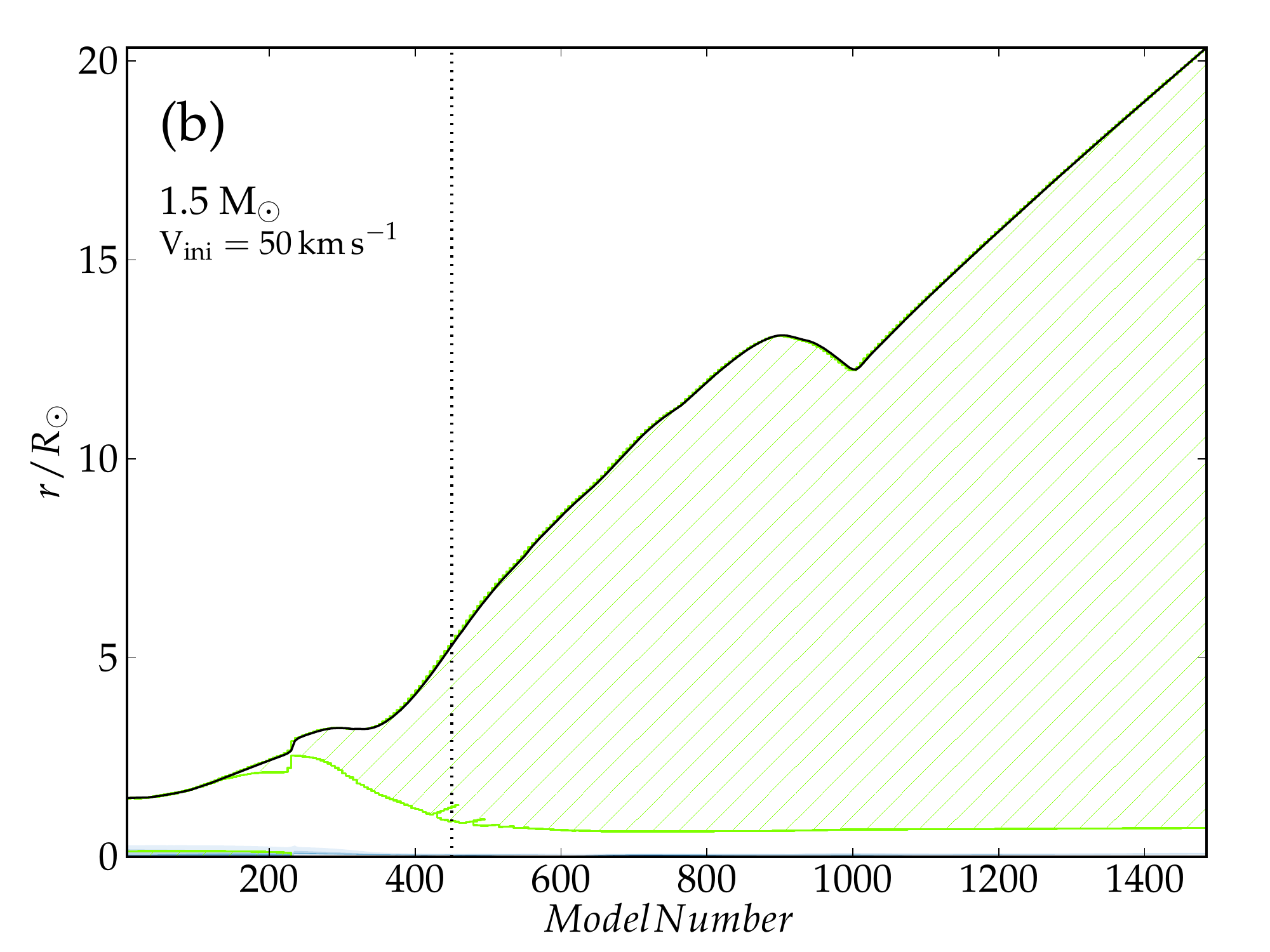}
\caption{Kippenhahn diagrams in mass (a) and radius (b) coordinate showing the evolution of a $1.5\mso$ model (as a function of model number) from the ZAMS to the RGB phase. Green hatched: convective,  blue shading: energy generation rate (minus neutrino losses). The location where we extract the background structures to calculate the splittings using ADIPLS (see Fig.~\ref{kernels} and \ref{splitting}) is identified by matching the observed values of $\nu_{\rm max}$ and $\Delta \nu$ and is  shown by the vertical dotted line. The luminosity bump occurs around model 1000. The evolution is shown up to $\llso\simeq2$ and $\teff \simeq 3980 K$. \label{kipp}}
\end{center}
\end{figure}

Convective regions are calculated using the mixing-length theory (MLT) in the \citet{Henyey:1965} formulation with $\alphaMLT=1.6$. Transport of angular momentum in convective regions is accounted for using the resulting MLT diffusion coefficient (turbulent diffusivity), which is so large as to cause rigid rotation in convective zones. While this seems to be the case in the Sun, another possible treatment of rotating convective zones is adopting a constant specific angular momentum \citep[See e.g.][]{Kawaler:2005}. We ran calculations with this assumption and found that it does not affect our conclusions. The boundaries of convective regions are determined using the Ledoux criterion. Semiconvection is accounted for in the prescription of \citet{Langer:1983,Langer:1985} with an efficiency $\alphasc$ = 0.003. 
A step function overshooting extends the mixing region for 0.2 pressure scale heights 
 beyond the convective boundary during core H-burning. 
We also account for gravitational settling and chemical diffusion \citep{Paxton:2011}. Figure~\ref{kipp} shows Kippenhahn diagrams for the $1.5\mso$ model.

\begin{figure}[h!]
\begin{center}
\includegraphics[width=1\columnwidth]{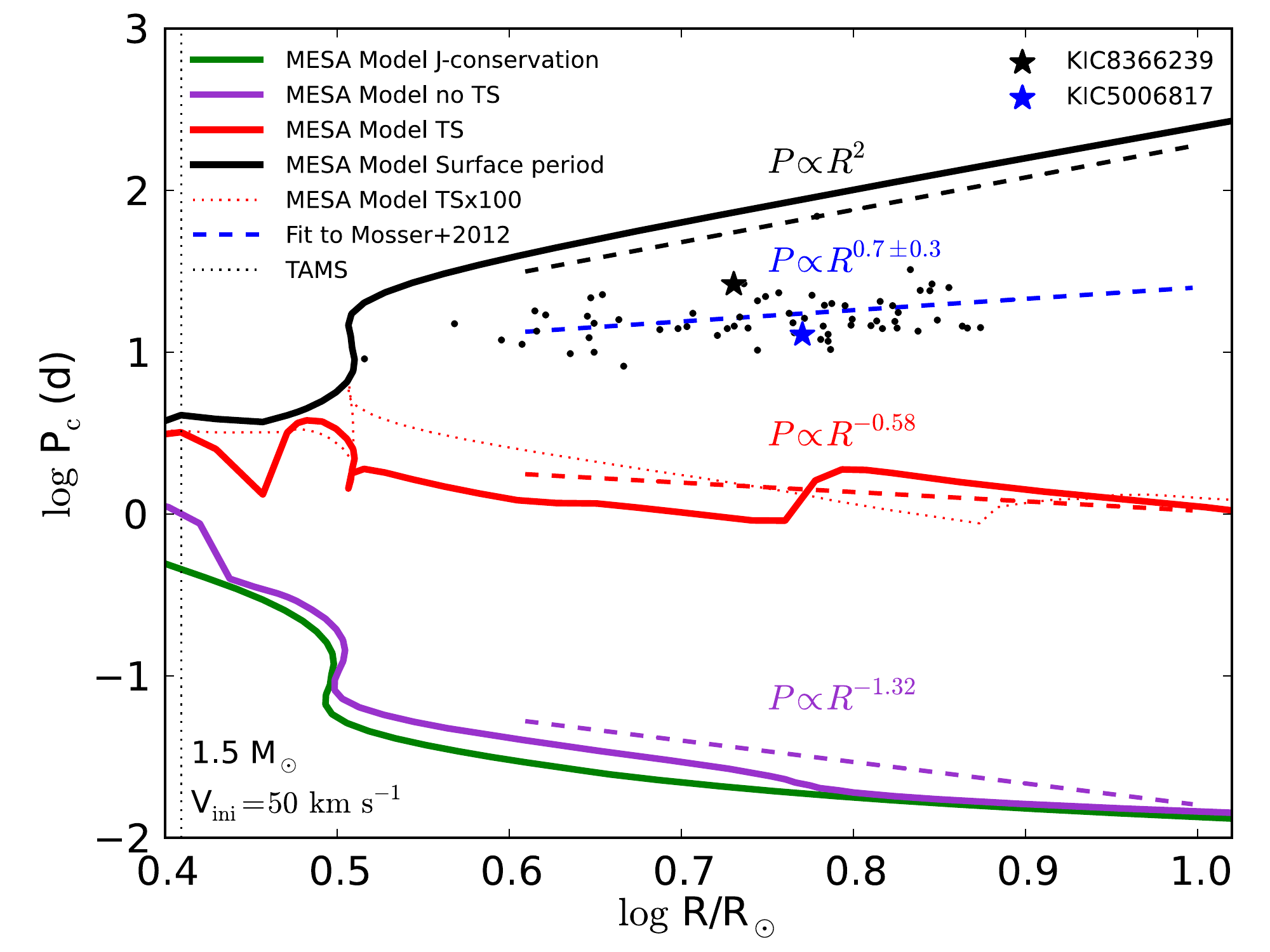}
\caption{Evolution of the average core rotational period as a function of stellar radius for different assumptions of angular momentum transport in a 1.5$\mso$ model initially rotating at 50$\kms$. We show models without angular momentum transport (green), including transport of angular momentum due to rotational instabilities (purple) and accounting for magnetic torques in radiative regions (red, Tayler-Spruit magnetic fields). The star symbols indicate the locations of $\KIC$ and $\KID$ as derived using the maximum observed splitting of their mixed modes \citep{Beck:2012,Beck:2014}. Dashed lines indicate a linear fit to the different curves during the early RGB. The vertical dotted line shows the location of H-core exhaustion. The red dotted line shows the evolution of core rotational period for a model where the resulting Tayler-Spruit diffusion coefficient has been multiplied by a factor of 100. Stars in the red giant sample of  \citet{Mosser:2012}  with $R<7.5\rso$ are shown as black dots. The best fit to the core rotation of the  \citet{Mosser:2012} sample is also shown as a dashed blue line.  \label{period}}
\end{center}
\end{figure}

\subsection{Early RGB}\label{earlyrgb}
The evolution of core rotation period, $P_{\rm c}$, during the early RGB for our  $1.5\mso$ models with different assumptions for the internal angular momentum transport  is shown in  Fig.~\ref{period}. The models are initially rigidly rotating at the ZAMS with a surface velocity of $50\kms$,  a typical value for stars in this mass range \citep{Nielsen:2013}.  The value shown for $P_{\rm c}$ is a mass average of the rotational period in the region below the maximum of the energy generation $\epsilon_{\rm nuc}$ in the H-burning shell. 
The contraction of the core leads to different rates of spin up for the different angular momentum transport mechanisms considered \citep[See e.g.][]{Tayar:2013}. At the same time the expanding envelope slows down substantially, following the expected $P_{\rm rot} \propto R^2$ scaling from expansion and angular momentum conservation (no substantial mass is lost in this phase).
These calculated cores rotate about $10-10^3$ times faster than the values inferred by asteroseismology (Fig.~\ref{period}). During this phase the mass of the core increases only slightly. 
The work of \citet{Mosser:2012} revealed that  the cores of stars in the mass range 1.2--1.5$\mso$  spin down while ascending the early RGB as $P_{\rm c} \propto R^{0.7\pm0.3}$, while our stellar evolution calculations show spin up with different slopes ($P_{\rm c} \propto R^{-0.58}$ for models including TS and $P_{\rm c} \propto R^{-1.32}$ for models only including  angular momentum transport by rotational instabilities\footnote{These exponents have been calculated for the range R/$\rso$=[3,15]}), depending on the assumptions for angular momentum transport.
This  clearly shows that the amount of torque between the core and envelope during the RGB evolution is underestimated by the models.

We explored whether an increase in the efficiency of the TS mechanism could reconcile the models with the observations. However, even increasing the diffusion coefficient resulting from the magnetic torques by a factor 100 does not result in enough coupling to explain the observations.  
This is largely due to the self-regulating nature of the Tayler-Spruit dynamo. The poloidal component of the magnetic field $B_r$ is generated by the Tayler instability that occurs in the toroidal component $B_{\phi}$ of the field. However the toroidal component is amplified by the  differential rotation, which is in turn suppressed by the torque $\propto B_r B_{\phi}$. It is not too surprising then to observe that the system tends to relax around some  differential rotation state which depends only weakly  on increasing the efficiency of the Tayler-Spruit dynamo loop.

Finally we checked if an artificial constant diffusivity $\nu$ could explain the observations. We confirm the results of \citet{Eggenberger:2012}: in our $1.5\mso$ model initially rotating with $50\kms$ a diffusivity of $\nu=6\times10^4\cms$ matches the observed splittings of $\KIC$. However such a constant diffusivity is unphysical and fails to explain the later evolution and the observations of rotation rates in clump stars and WDs (see Sec.~\ref{clump}).

We conclude that on the early RGB of low-mass stars none of the angular momentum transport mechanisms usually included in stellar evolution codes can produce a  coupling adequate to explain the asteroseismic derived core rotation rates.

\section{Calculating the splitting of mixed modes}\label{splittings}
Rotation lifts the degeneracy between the non-radial modes of the same radial order $n$ and degree $\ell$ but different azimuthal order $m$. When the rotation of the star is slow  the centrifugal force can be neglected.  
If the rotation profile is spherically symmetric one obtains 
\begin{equation}
\nu_{n,\ell,m}=\nu_{n,\ell,0}+m\,\delta\nu_{n,\ell}\, ,
\end{equation} 
for the frequency of the $(n,\ell,m)$  mode. Here $\delta\nu_{n,\ell}$ is the rotational splitting, a weighted measure of the star's rotation profile $\Omega(r)$, given by
\begin{equation}
\delta\nu_{n,\ell} = \frac{1}{2\pi} \int_0^R K_{n,\ell}(r) \, \Omega(r) \,\hbox{d}r\, .
\end{equation}
The functions $K_{n,\ell}(r)$ are called rotational kernels of the modes and depend on the star's equilibrium structure and on the mode eigenfunctions \citep{Aerts:2010}. 
Therefore rotational splittings are a weighted measure of the star{'}s rotation rate through the
rotational kernels.

In order to calculate the splitting of mixed modes we used the adiabatic pulsation code ADIPLS \citep[][2011 June release]{JCD:2008}. This code is coupled and distributed within the MESA code suite \citep{Paxton:2013}.

We start by modeling the red giant $\KIC$, for which \citet{Beck:2012} observed a rotational splitting of p-dominated mixed modes of $\delta\nu_{n,1} = 0.135\pm0.008 \mu$Hz. On the other hand,  the observed splitting for the g-dominated mixed modes (mostly living in the stellar core) is $0.2-0.25\mu$Hz. This confirmed the theoretical expectation that the core of this red giant is rotating faster than its envelope. 
Similarly to \citet{Eggenberger:2012} for $\KIC$ we adopt an initial mass of $1.5\mso$ and calculate models assuming different physics for angular momentum transport. We select the background structure of the different calculations by matching the global asteroseismic properties of $\KIC$ (frequency of maximum oscillation power, $\nu_{\rm max}$, and large frequency separation $\Delta \nu$), as derived by \citet{Beck:2012}. The MESA background structure and rotational profiles are then used in ADIPLS to calculate the splitting of mixed modes for different  angular momentum transport mechanisms.  

\begin{figure}[ht!]
\begin{center}
\includegraphics[width=1\columnwidth]{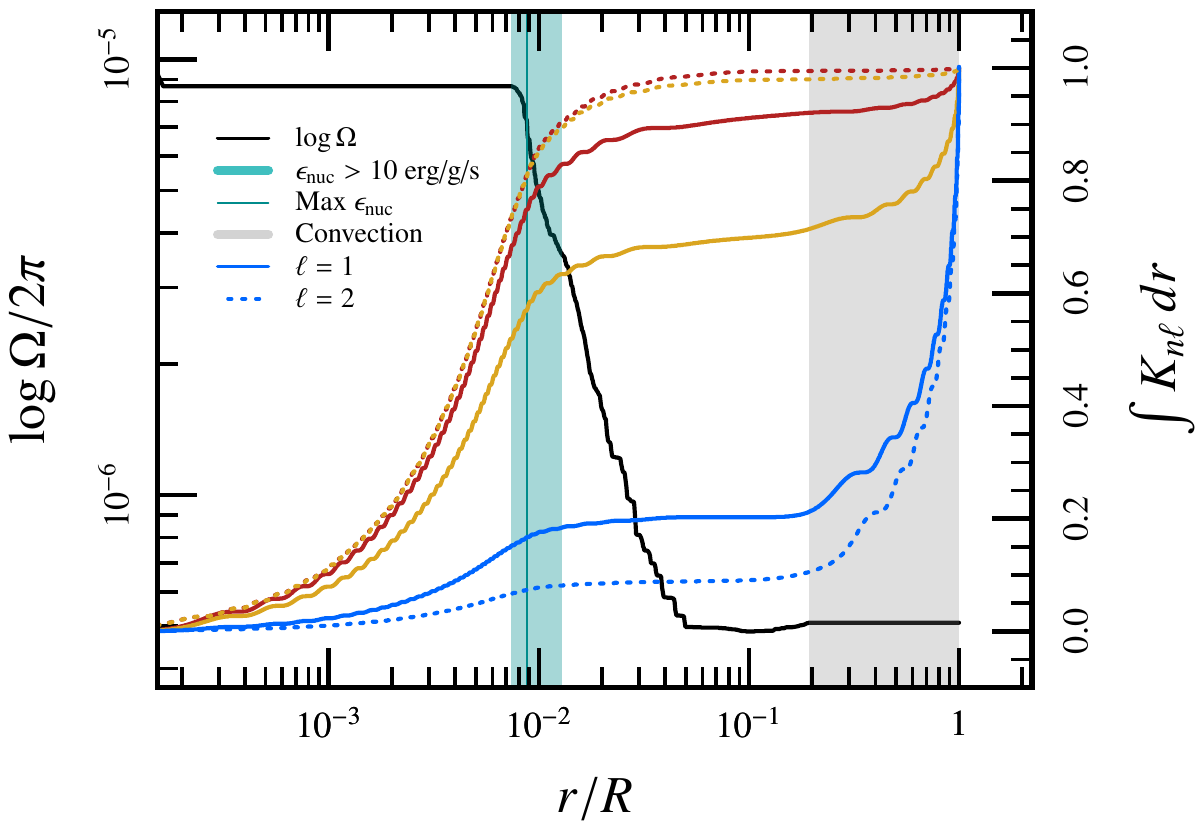}
\caption{ADIPLS calculations of the integrated kernels for g-dominated (yellow and red) and p-dominated (blue) mixed modes. The background model has an initial mass of $1.5\mso$, an initial equatorial rotational velocity of $150\kms$ and has been evolved including the effect of magnetic torques from Tayler-Spruit dynamo fields. The x-axis is showing the normalized radial coordinate (the model has a radius of about $5.4\rso$, $\log\teff \simeq 3.67$ and $\llso \simeq 1.21$). Solid lines are for $\ell=1$ and dashed lines for $\ell=2$  (right-hand ordinate scale) The x-axis is showing the normalized radial coordinate. Solid lines are for $\ell=1$ and dashed lines for $\ell=2$  (right-hand ordinate scale). The black solid line represents the angular velocity profile of the background structure model calculated with MESA  (left-hand ordinate scale). The convective envelope is shown in gray, while the location of the H-burning shell is marked in turquoise\label{kernels}.   }
\end{center}
\end{figure}

We show in Fig.~\ref{kernels} an example of the background rotational profile calculated by MESA, together with the radial partial integrals of the rotational kernels $K_{n,\ell}$ for $\ell=1,2$ calculated using ADIPLS. This  shows that the p-dominated modes mostly probe the envelope of the star, where the angular velocity is quite low. The gravity dominated modes  probe the radiative region below the H-burning shell, in which the angular velocity is much higher. Note that higher $\ell$ modes have higher Lamb frequencies, implying a larger tunneling zone between the acoustic cavity in the envelope and the gravity mode region in the core. As a consequence  these modes are ``less mixed'', with $\ell=2$ p-modes (g-modes) being more p-like (g-like) than their $\ell=1$ analogue. Therefore when observable, higher $\ell$ g-dominated (p-dominated) modes are cleaner probes of core (envelope) rotation rate.

\begin{figure}[ht!]
\begin{center}
\includegraphics[width=1\columnwidth]{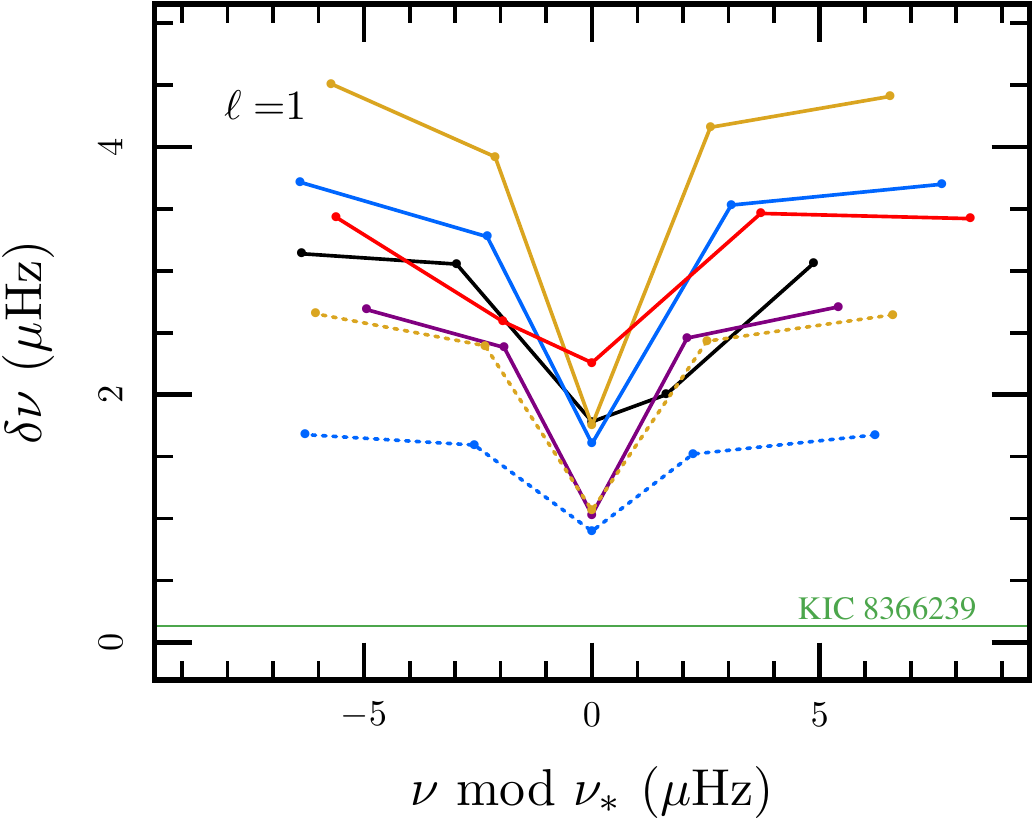}
\includegraphics[width=1\columnwidth]{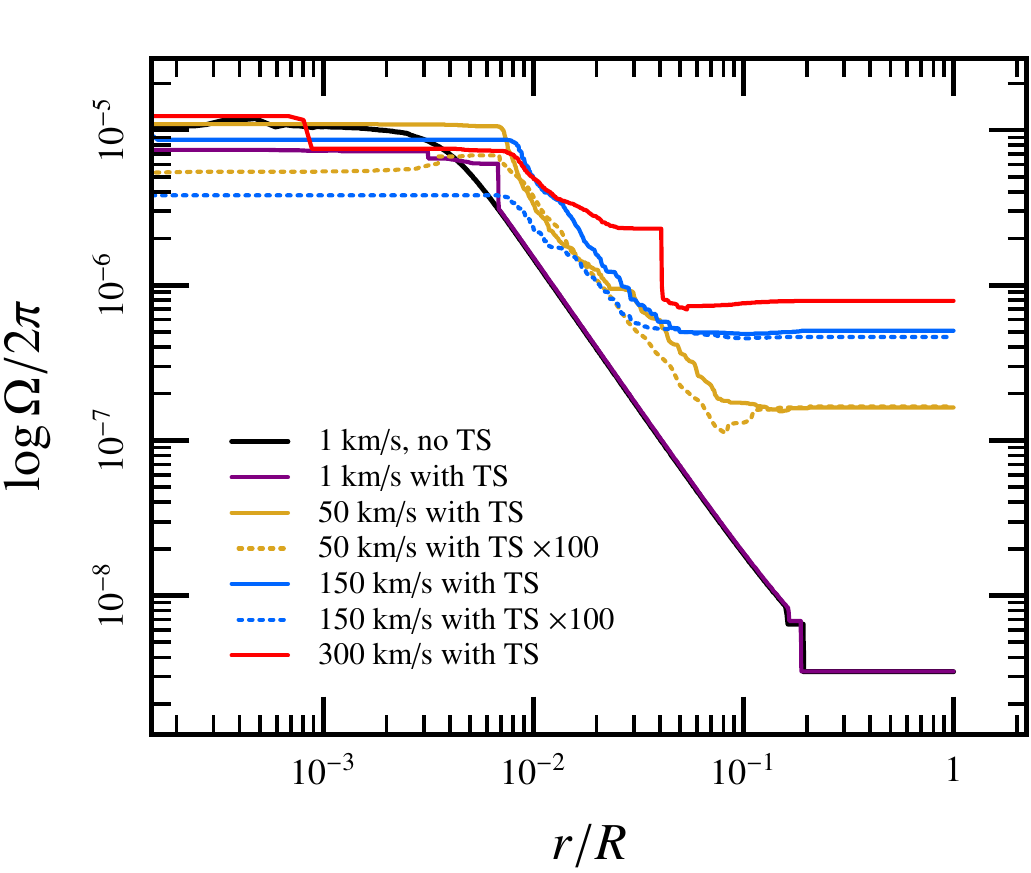}
\caption{Calculated rotational splittings $\delta\nu$ for $\ell=1$ mixed modes near $\nu_*$, the frequency of the p-dominated mode closest to $\nu_{\rm max}\approx190 \mu$Hz (top panel). The splittings are calculated from our $1.5\mso$ models for different assumptions of angular momentum transport and initial rotational velocities. The minimum splitting of $\delta\nu_{n,1} = 0.135 \mu$Hz observed for $\KIC$ is shown by the horizontal line for reference.  The corresponding angular velocity profiles are shown in the bottom panel as a function of normalized radial coordinate. Models where the diffusion of angular momentum from Tayler-Spruit dynamo fields has been artificially enhanced by a factor of 100 are shown as dashed lines. \label{splitting}}
\end{center}
\end{figure}

To carefully compare the model rotation rates with those derived from the observations we calculate the relevant eigenfunctions and rotational kernels $K_{n,\ell}(r)$ and compute the predicted splittings of $\KIC$ as a function of different initial rotational velocities and physics of angular momentum transport (Fig.~\ref{splitting}). The splittings have been calculated using the ADIPLS code and the MESA background structure at the location that matches the asteroseismic properties ($\nu_{\rm max}$, $\Delta \nu$) of the star.

In agreement with results of Sec.~\ref{earlyrgb}, models that only include angular momentum transport due to rotational instabilities and circulations fail to reproduce the observed splittings in the RGB star $\KIC$. 
Even models with an extremely slow initial rotation of $1 \kms$ result in rotational splittings one order of magnitude larger than the observed ones, which clearly shows this class of models can not explain the observations.
This agrees with the calculations of \citet{Eggenberger:2012}, despite the implementation of the physics of rotation being quite different in the GENEVA code compared to MESA \citep[See Sec. 6 in ][and references therein]{Paxton:2013}.

Models including angular momentum transport due to Tayler-Spruit magnetic fields  couple much more strongly, but still result in rotational splittings on the order of 1~$\mu$Hz (Fig.~\ref{splitting}), a factor of $\ga10$ higher than measured.
Even artificially increasing the Tayler-Spruit diffusion coefficient by a factor of 100 we could not reproduce the observed splittings.

\begin{figure}[h!]
\begin{center}
\includegraphics[width=1\columnwidth]{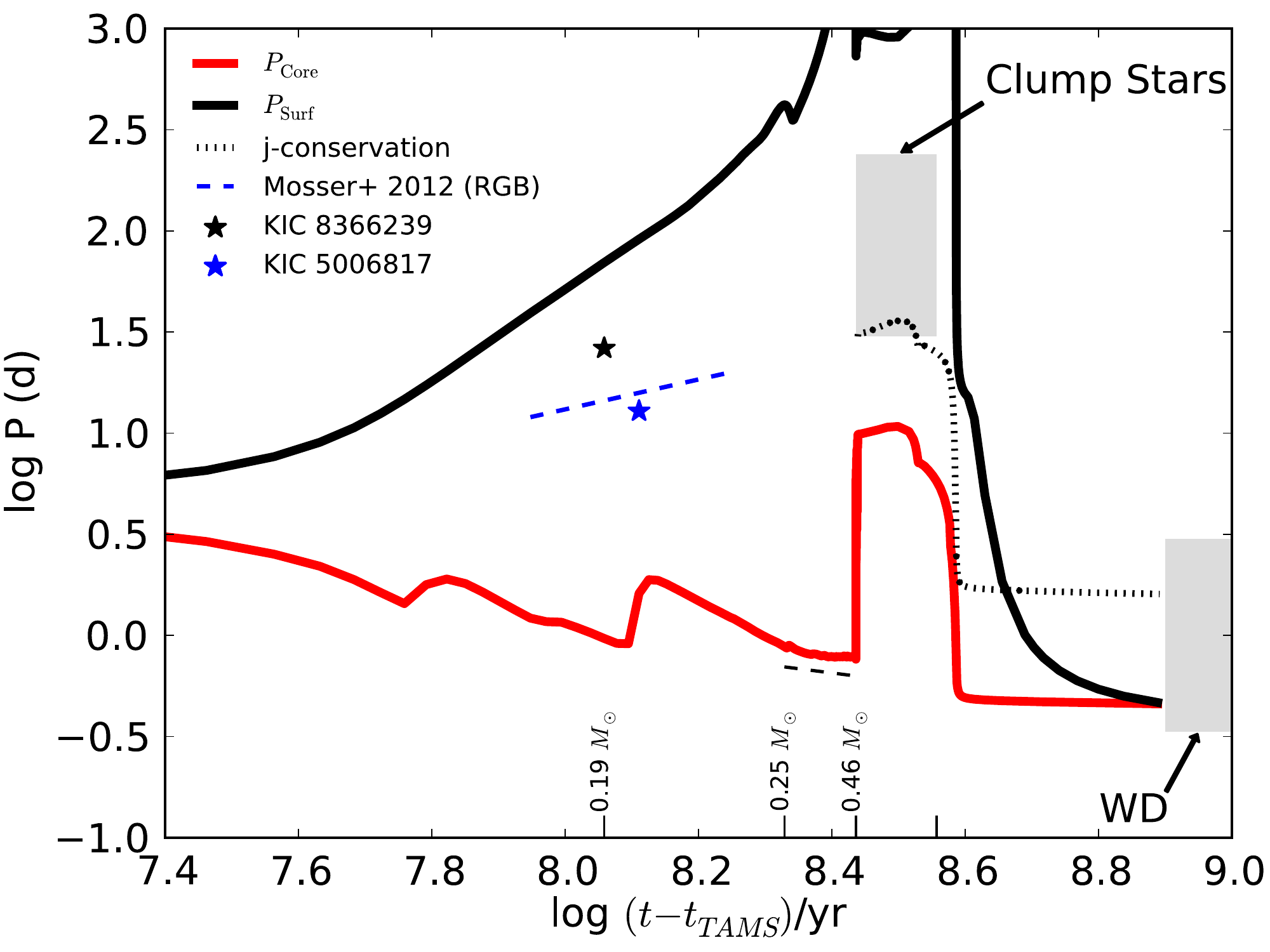}
\caption{Evolution of the average core rotational period as a function of logarithmic time to the end of the calculation (WD cooling sequence). The black and red solid lines show the surface and core rotational periods for our 1.5$\mso$ model rotating with an initial surface velocity of about 50$\kms$ and including angular momentum transport due to Tayler-Spruit magnetic fields in radiative regions (together with transport due to rotational instabilities). The plot shows evolution from the end of H-core burning (terminal age main sequence, TAMS) to the cooling WD sequence (when the luminosity decreases below 
$\log L/\lso \sim$ {-}2).	The	star symbols indicate the location	of $\KIC$ and $\KID$ as derived using the maximum observed splitting of their mixed modes, the blue dashed line shows the fit to the core rotation data of early red giants in \citet{Mosser:2012}. Gray regions show the range of the inferred core rotation rates for the clump stars of \citet{Mosser:2012} and the typical rotation rate of non magnetic WD \citep[see e.g.][]{Koester:1998,Ferrario:2005,Corsico:2011,Greiss:2014}. The dotted line shows a model with an imposed core rotation of 30 d on the clump evolving with no angular momentum transport ($j$-conservation) into a WD.  The black dashed line shows a fit to the theoretical prediction for core rotation on the RGB past the luminosity bump, highlighting the change in slope compared to the early RGB (see text). \label{period_evolution}}
\end{center}
\end{figure}

\section{Angular momentum evolution beyond the RGB}\label{beyond}

\subsection{RGB past the luminosity bump}\label{bump}
As the degenerate He core grows in mass, its angular momentum content is determined by the rate of angular momentum transport  and by the specific angular momentum of the advected material. Both may change as the star climbs up the RGB. This is because angular momentum transport mechanisms can be a function of, e.g., the rate of shear between the core and the envelope, which tends to increase as the star expands. 
Moreover during the RGB the H-burning shell moves up in mass coordinate and at some point crosses the  compositional discontinuity left by the first dredge up (luminosity bump).
Since the envelope expands and at the same time loses a considerable amount of mass through stellar winds (about $0.3\mso$ in the $1.5\mso$ model), it loses angular momentum at an increasing rate. As angular momentum is expected to be mixed efficiently in convective regions, the specific angular momentum of the material engulfed by the core after the luminosity bump is expected to be low and to decrease as the star climbs the RGB. Note that the disappearance of the steep compositional gradient after the luminosity bump is also expected to enhance the efficiency of angular momentum transport mechanisms between core and envelope. Evidence of enhanced chemical mixing below the convective envelope (cold bottom process) comes from the observation of surface abundances in red giants, in particular a sudden drop in the carbon isotopic ratio $^{12}{\rm C}/^{13}{\rm C}$ and changes in $^7{\rm Li}$, carbon and nitrogen \citep{Gratton:2000}. The nature of this mixing is currently debated \citep[See e.g.][]{Palacios:2006,Charbonnel:2007,Nordhaus:2008,Cantiello:2010,Traxler:2011,Denissenkov:2011,Brown:2013}.
   
In our $1.5\mso$ calculations the luminosity bump occurs when the star has a value of the large separation $\Delta\nu \simeq 3.6\mu$Hz and frequency of maximum oscillation power $\nu_{\rm max}\simeq 30.5\mu$Hz. Regardless of the specific angular momentum transport mechanism included, we find a change in the exponent of the $P_c\propto R^{\,\xi}$ relation associated with the luminosity bump (see e.g. the black dashed line in Fig.~\ref{period_evolution}). In particular for the model including magnetic torques and rotating with an initial surface velocity of $50\kms$, $\xi$ changes from -0.58 to -0.01 (while the same model only including angular momentum transport due to rotational instabilities has $\xi$ changing from -1.32 to -0.13). Different exponents are found for different initial rotational velocities, but we consistently find a break at the luminosity bump.
This is because the value of the specific angular momentum of the advected material decreases rapidly as the core engulfs regions left by the retreating convective envelope.
Therefore, regardless of the specific angular momentum transport mechanism operating in stars, in red giants ascending the RGB we expect that the rate of spindown should decrease past the luminosity bump, and depart from the relation $\approx R^{0.7\pm0.3}$ observed by \citet{Mosser:2012}.

While such a change in the exponent could give further clues into the currently debated extra mixing mechanism that operate past the luminosity bump, it is unlikely to be observed, as g-dominated modes are predicted to become unobservable as stars move up along the RGB, due to a combination of increasing inertia and increasing damping in the core \citep{Dupret:2009}. For a $1.5\mso$ and 1 year of observations g-dominated mixed modes have been predicted to be detectable only for stars with $\nu_{\rm max} \ge 50\mu$Hz and $\Delta\nu \ge 4.9\mu$Hz \citep{Grosjean:2013}.

\subsection{Clump stars}\label{clump}
After reaching the tip of the RGB, stars with $M\lesssim 2\mso$ ignite He in their degenerate core. This leads to a large release of energy, called the He-flash, which during a period of about 2 Myr lifts the degeneracy of the core leading to a stable He-burning phase. 
Such a transition phase has a unique asteroseismic signature \citep{2012ApJ...744L...6B}.
 
In our models during the He-flash the rotational period of the core increases quite rapidly by a factor of about 10. This is because the nuclear energy released results in core expansion. In our $1.5\mso$ model the $0.46 \mso$ core expands by approximately a factor of 3 during the He-flash, with core moment of inertia increasing by a factor of 10 from $I_{\rm c}=3.13\times10^{50}\, \GramSc$ to $I_{\rm c}=3.06\times10^{51} \,\GramSc$ \citep[See also][]{Kawaler:2005}, fully accounting for the spin down observed in the models (See Fig.~\ref{period_evolution}).
Even if the timescale of the He-flash is too short for angular momentum transport outside the core, we note that the convective episodes that accompany the He flash can potentially play an important role in the redistribution of angular momentum inside the He-core.
Such rapid mixing episodes can change the rotational profile of the g-mode cavity, as they lead to a fairly rigidly rotating radiative region above core He-burning. Therefore the expectation is that, regardless of previous history of angular momentum transport, the core of clump stars that underwent ignition of He in a degenerate core should be nearly rigidly-rotating.

After this rapid initial phase, the core rotation rate remains fairly constant during core He burning.  
The clump stars in the \citet{Mosser:2012} sample rotate  with periods in the range $P_{\rm c} \sim 30-240$ d. Isolated pulsating sdB stars  (red giants stripped of their envelope) show similar rotation rates,  with periods ranging from 23 to 88 d \citep[See e.g.][]{Baran:2012}. Similar to the case of the early RGB, these values are about 1 order of magnitude slower than models which include magnetic torques, again pointing toward the need for some extra angular momentum transport occurring in previous evolutionary phases.
Note that models including an artificial diffusivity able to reproduce the observed splitting on the early RGB ($\nu \sim 10^4-10^5 \cms$) fail to explain the rotation rates of clump stars, with predicted rotation rates almost two orders of magnitude higher than the observations. This is because the torque required to couple core and envelope increases as the star rapidly climbs the RGB. 
We note that a combination of an artificial viscosity $\nu \sim 10^4-10^5 \cms$ with the Tayler-Spruit magnetic torques can reproduce both the early RGB and the clump observations.

\subsection{White Dwarf Rotation Rates}\label{wd}
After core He burning the energy generation proceeds in two shells (burning H and He) moving outwards in mass coordinate while the star moves up the asymptotic giant branch (AGB). The He-shell becomes secularly unstable, giving rise to thermal pulses (TP-AGB). These pulses grow in intensity and are thought to enhance mass loss, ultimately leading to a complete removal of the H-envelope, a planetary nebula and the transition to the white dwarf cooling sequence. The details are not well understood and the transition time from the AGB to the WD cooling sequence depends on the treatment of mass-loss beyond the AGB. Note however that the timescale for angular momentum transport between core and envelope is likely much longer than the range of timescales discussed for the duration of this phase, so that the results on the angular momentum content of WD models should not depend strongly on the particular treatment of this phase. This is supported by the fact that the observed WD rotation rates can be recovered from the observed core rotation of clump stars assuming no angular momentum transport (see dotted line in Fig.~\ref{period_evolution}). It is true that this is only achieved assuming clump rotation rates on the high end of the observed distribution; however one has to be careful  as the mass range of the observed clump stars does not necessarily match the mass range of the progenitors of the WD with observed rotation rates.

\begin{figure}[ht!]
\begin{center}
\includegraphics[width=1\columnwidth]{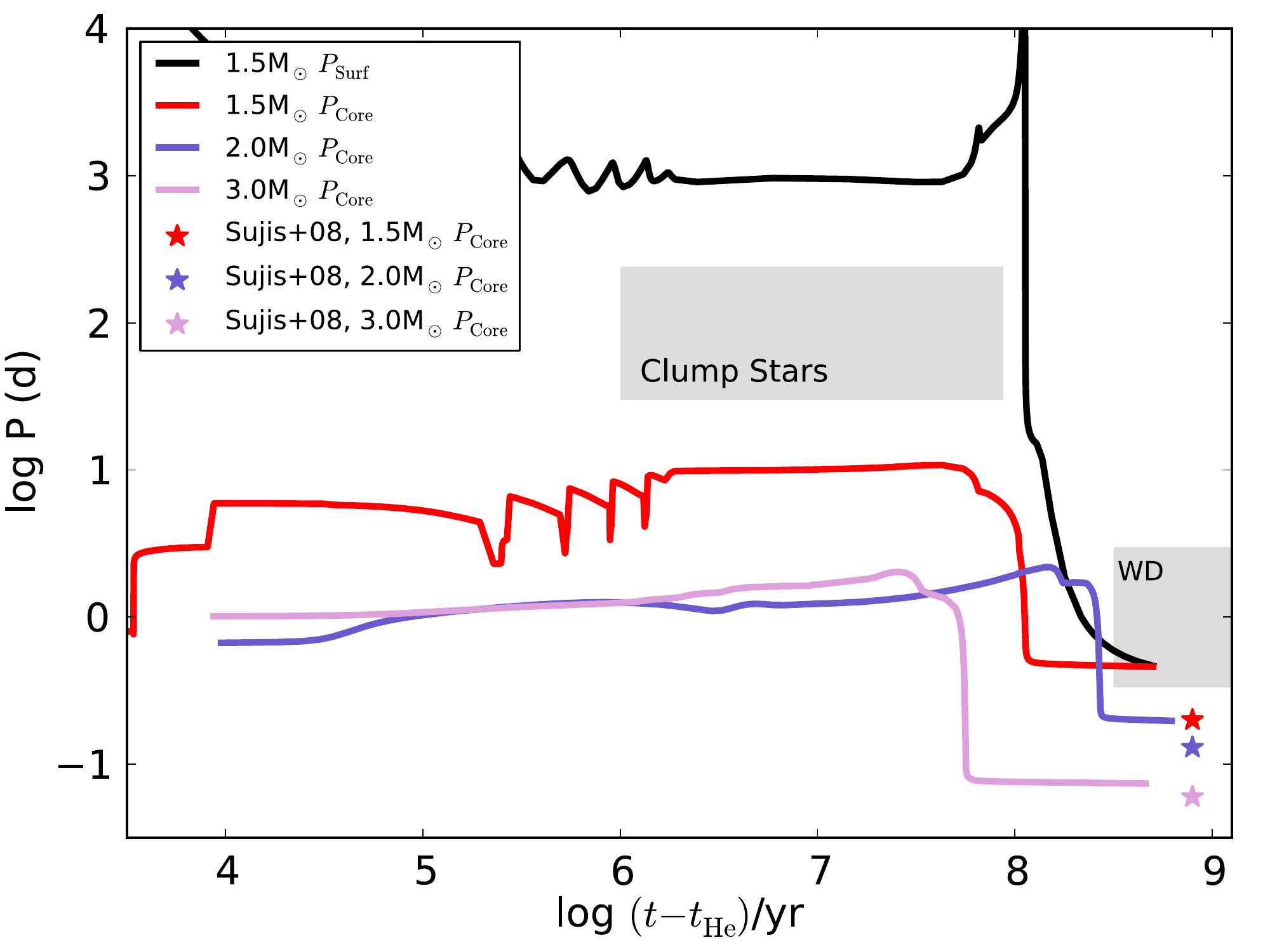}
\caption{Like Fig.~\ref{period_evolution} but showing only the evolution past He ignition. Different lines show the core rotation rate for 1.5, 2.0 and 3.0 $\mso$ models initially rotating with 50, 50 and 150 $\kms$ respectively. All models include Tayler-Spruit magnetic fields. The black line shows the surface rotational period of the 1.5$\mso$ model. Final rotational period for the 1.5, 2.0 and 3.0$\mso$ of \citet{larends_Yoon_Heger_Herwig_2008} are shown by star symbols. These models have also been calculated including magnetic torques but with initial rotational velocities 45, 140 and 250 $\kms$ respectively. The 1.5$\mso$ model ignites He in a degenerate core, resulting in a $\sim 2 $ Myr phase in which the degeneracy is removed through repeated convective events (He-core flash). \label{pzoom}}
\end{center}
\end{figure}

In our calculations during the TP-AGB phase we adopt the mass loss prescription of \citet{Bloecker:1995} multiplied by an efficiency factor  $\eta = 10$. Results for the final rotation rate of WD as predicted by our models including magnetic torques are shown in Fig.~\ref{pzoom}. We note that, while they fail to explain the rotation rates in previous evolutionary phases (RGB and clump), these models can marginally reproduce values deduced from the asteroseismic observations of ZZ Ceti stars (pulsating WDs). This is consistent with the findings of \citet{larends_Yoon_Heger_Herwig_2008}, even if their final rotation rates seem to be slightly higher than the one found by the MESA calculations (which are in better agreement with the observations).

\section{Conclusions and future work}\label{conclusions}
We found that stellar evolution calculations of low-mass stars, including angular momentum transport from 
rotational instabilities, rotationally induced circulations and magnetic torques from Tayler-Spruit dynamo fields, can not explain the observed core rotation of early red giants and clump stars. 
The physics of internal angular momentum transport in stars is still not understood.
The asteroseismic observations imply cores that are rotating at least 10 times slower than the predicted values. This is in much better agreement than the result of models only including angular momentum transport by rotational instabilities  and  circulations, which predict cores rotating more than $10^3$ times faster than the observed values. 
By considering the later stages of stellar evolution, we identify the RGB as the evolutionary phase where some extra angular momentum transport mechanism is efficiently coupling the core to the stellar envelope. Potential candidates for such mechanism are internal gravity waves and fossil (or convective dynamo-generated) magnetic fields. 

Ensemble asteroseismology is providing outstanding results regarding the internal rotation of low-mass stars. 
It is important, however, to keep in mind that only a fraction of the stars analyzed for seismology have identifiable mixed modes and rotational splittings  \citep[in the case of][for example, 313 out of 1399 stars, i.e. 22\%]{Mosser:2012}.
 If the core rotates rapidly rotational multiplets are highly complex   \citep{Ouazzani:2013}; this could in principle have led to a bias toward stars with slower rotating cores (e.g. stars that did spin down due to binary interactions or because of a fossil magnetic field). We believe these potential biases need to be carefully addressed as they could have important repercussion on the theoretical interpretations. 

Nevertheless it is interesting to consider what angular momentum transport mechanisms could be responsible for the strong coupling implied by the asteroseismic observations. One candidate are gravity waves excited by the convective envelope during the RGB, as these can potentially lead  to some transport of angular momentum. While a similar process has been discussed in the context of the Sun's rotational profile \citep{Zahn:1997,Charbonnel:2005}, more work needs to be done to understand the details of the excitation and propagation of gravity waves \citep[See e.g.][]{Lecoanet:2013,Shiode:2013,Rogers:2013}

Another possibility is that some large scale magnetic field is present in and above the stellar core at the end of the main sequence, providing some coupling between core and envelope. This magnetic field could be either of fossil origin (similar to what has been discussed in the context of explaining the internal rotation profile of the Sun) or be generated by a convective dynamo in the H-burning core during the main sequence. Dynamo action is favorable as, given the typical rotational velocities of 1.5-3.0$\mso$ stars during the main sequence, Rossby numbers are usually smaller than 1, implying an $\alpha\Omega$-dynamo could be at work in the core. The equipartition magnetic field is  $B_{\rm{eq}} = \varv_c\,\sqrt{4\pi\rho}$; assuming $B_{\phi}\sim B_r\sim B_{\rm{eq}}$ the resulting magnetic stress is $S= B_r B_{\phi}/4\pi$ and the associated diffusivity is 
 $\nu \sim S/(\rho q \Omega)$, where $q=-\partial \log \Omega/\partial \log r$ is the shear. Typical convective velocities in the core of main-sequence, low-mass stars are on the order $0.01\kms$ resulting in $B_{\rm{eq}}\sim 10^4-10^5 G$. Some of the magnetic flux will diffuse in the radiative layers above the convective core, but this is expected to affect only a small fraction of the star as the Ohmic diffusion timescale is much longer than the main sequence timescale. 
Overall assessing whether such a mechanism can explain the observed rotation rates requires following the coupled evolution of shear and magnetic fields.
 
We thank the anonymous referee for their comments and suggestions, which contributed to improve this paper. 
We thanks Conny Aerts, Steve Kawaler,  Sterl Phinney, Marc Pinsonneault, Eliot Quataert, Dennis Stello, Rich Townsend and Jim Fuller for helpful discussions. 
We are grateful to Beno\^{i}t Mosser for sharing his data.
This project was supported by
NASA under TCAN grant number NNX14AB53G and
the NSF under grants PHY 11-25915 and AST 11-09174. 
Funding for the Stellar Astrophysics Centre is provided by 
The Danish National Research Foundation (Grant DNRF106). The research is supported by the ASTERISK project (ASTERoseismic Investigations with SONG and \textit{Kepler}) funded by the European Research Council (Grant agreement no.: 267864).

\vspace{0.2ex} 
The inlist adopted  to calculate the models in this paper can be found at \url{https://authorea.com/1608/} (bottom of the document).


\begin{thebibliography}{}
\expandafter\ifx\csname natexlab\endcsname\relax\def\natexlab#1{#1}\fi

\bibitem[{{Aerts} {et~al.}(2010){Aerts}, {Christensen-Dalsgaard}, \&
  {Kurtz}}]{Aerts:2010}
{Aerts}, C., {Christensen-Dalsgaard}, J., \& {Kurtz}, D.~W. 2010,
  {Asteroseismology} (Springer)

\bibitem[{{Asplund} {et~al.}(2005){Asplund}, {Grevesse}, \&
  {Sauval}}]{Asplund:2005}
{Asplund}, M., {Grevesse}, N., \& {Sauval}, A.~J. 2005, in Astronomical Society
  of the Pacific Conference Series, Vol. 336, Cosmic Abundances as Records of
  Stellar Evolution and Nucleosynthesis, ed. T.~G. {Barnes}, III \& F.~N.
  {Bash}, 25

\bibitem[{{Baran} {et~al.}(2012){Baran}, {Reed}, {Stello}, {{\O}stensen},
  {Telting}, {Pak{\v s}tien{\"e}}, {O'Toole}, {Silvotti}, {Degroote},
  {Bloemen}, {Hu}, {Van Grootel}, {Clarke}, {Van Cleve}, {Thompson}, \&
  {Kawaler}}]{Baran:2012}
{Baran}, A.~S., {Reed}, M.~D., {Stello}, D., {et~al.} 2012, \mnras, 424, 2686

\bibitem[{{Beck} {et~al.}(2011){Beck}, {Bedding}, {Mosser}, {Stello}, {Garcia},
  {Kallinger}, {Hekker}, {Elsworth}, {Frandsen}, {Carrier}, {De Ridder},
  {Aerts}, {White}, {Huber}, {Dupret}, {Montalb{\'a}n}, {Miglio}, {Noels},
  {Chaplin}, {Kjeldsen}, {Christensen-Dalsgaard}, {Gilliland}, {Brown},
  {Kawaler}, {Mathur}, \& {Jenkins}}]{Beck:2011}
{Beck}, P.~G., {Bedding}, T.~R., {Mosser}, B., {et~al.} 2011, Science, 332, 205

\bibitem[{{Beck} {et~al.}(2012){Beck}, {Montalban}, {Kallinger}, {De Ridder},
  {Aerts}, {Garc{\'{\i}}a}, {Hekker}, {Dupret}, {Mosser}, {Eggenberger},
  {Stello}, {Elsworth}, {Frandsen}, {Carrier}, {Hillen}, {Gruberbauer},
  {Christensen-Dalsgaard}, {Miglio}, {Valentini}, {Bedding}, {Kjeldsen},
  {Girouard}, {Hall}, \& {Ibrahim}}]{Beck:2012}
{Beck}, P.~G., {Montalban}, J., {Kallinger}, T., {et~al.} 2012, \nat, 481, 55

\bibitem[{{Beck} {et~al.}(2014){Beck}, {Hambleton}, {Vos}, {Kallinger},
  {Bloemen}, {Tkachenko}, {Garc{\'{\i}}a}, {{\O}stensen}, {Aerts}, {Kurtz}, {De
  Ridder}, {Hekker}, {Pavlovski}, {Mathur}, {De Smedt}, {Derekas}, {Corsaro},
  {Mosser}, {Van Winckel}, {Huber}, {Degroote}, {Davies}, {Pr{\v s}a},
  {Debosscher}, {Elsworth}, {Nemeth}, {Siess}, {Schmid}, {P{\'a}pics}, {de
  Vries}, {van Marle}, {Marcos-Arenal}, \& {Lobel}}]{Beck:2014}
{Beck}, P.~G., {Hambleton}, K., {Vos}, J., {et~al.} 2014, \aap, 564, A36

\bibitem[{{Bildsten} {et~al.}(2012){Bildsten}, {Paxton}, {Moore}, \&
  {Macias}}]{2012ApJ...744L...6B}
{Bildsten}, L., {Paxton}, B., {Moore}, K., \& {Macias}, P.~J. 2012, \apjl, 744,
  L6

\bibitem[{{Bloecker}(1995)}]{Bloecker:1995}
{Bloecker}, T. 1995, \aap, 297, 727

\bibitem[{{Braithwaite}(2006)}]{Braithwaite:2006}
{Braithwaite}, J. 2006, \aap, 449, 451

\bibitem[{{Brown} {et~al.}(2013){Brown}, {Garaud}, \& {Stellmach}}]{Brown:2013}
{Brown}, J.~M., {Garaud}, P., \& {Stellmach}, S. 2013, \apj, 768, 34

\bibitem[{{Cantiello} \& {Langer}(2010)}]{Cantiello:2010}
{Cantiello}, M., \& {Langer}, N. 2010, \aap, 521, A9

\bibitem[{{Ceillier} {et~al.}(2013){Ceillier}, {Eggenberger}, {Garc{\'{\i}}a},
  \& {Mathis}}]{Ceillier:2013}
{Ceillier}, T., {Eggenberger}, P., {Garc{\'{\i}}a}, R.~A., \& {Mathis}, S.
  2013, \aap, 555, A54

\bibitem[{{Charbonnel} \& {Talon}(2005)}]{Charbonnel:2005}
{Charbonnel}, C., \& {Talon}, S. 2005, Science, 309, 2189

\bibitem[{{Charbonnel} \& {Zahn}(2007)}]{Charbonnel:2007}
{Charbonnel}, C., \& {Zahn}, J.-P. 2007, \aap, 467, L15

\bibitem[{{Christensen-Dalsgaard}(2008)}]{JCD:2008}
{Christensen-Dalsgaard}, J. 2008, \apss, 316, 113

\bibitem[{{C{\'o}rsico} {et~al.}(2011){C{\'o}rsico}, {Althaus}, {Kawaler},
  {Miller Bertolami}, {Garc{\'{\i}}a-Berro}, \& {Kepler}}]{Corsico:2011}
{C{\'o}rsico}, A.~H., {Althaus}, L.~G., {Kawaler}, S.~D., {et~al.} 2011,
  \mnras, 418, 2519

\bibitem[{{Deheuvels} {et~al.}(2012){Deheuvels}, {Garc{\'{\i}}a}, {Chaplin},
  {Basu}, {Antia}, {Appourchaux}, {Benomar}, {Davies}, {Elsworth}, {Gizon},
  {Goupil}, {Reese}, {Regulo}, {Schou}, {Stahn}, {Casagrande},
  {Christensen-Dalsgaard}, {Fischer}, {Hekker}, {Kjeldsen}, {Mathur}, {Mosser},
  {Pinsonneault}, {Valenti}, {Christiansen}, {Kinemuchi}, \&
  {Mullally}}]{Deheuvels:2012}
{Deheuvels}, S., {Garc{\'{\i}}a}, R.~A., {Chaplin}, W.~J., {et~al.} 2012, \apj,
  756, 19

\bibitem[{{Deheuvels} {et~al.}(2014){Deheuvels}, {Do{\u g}an}, {Goupil},
  {Appourchaux}, {Benomar}, {Bruntt}, {Campante}, {Casagrande}, {Ceillier},
  {Davies}, {De Cat}, {Fu}, {Garc{\'{\i}}a}, {Lobel}, {Mosser}, {Reese},
  {Regulo}, {Schou}, {Stahn}, {Thygesen}, {Yang}, {Chaplin},
  {Christensen-Dalsgaard}, {Eggenberger}, {Gizon}, {Mathis},
  {Molenda-{\.Z}akowicz}, \& {Pinsonneault}}]{Deheuvels:2014}
{Deheuvels}, S., {Do{\u g}an}, G., {Goupil}, M.~J., {et~al.} 2014, \aap, 564,
  A27

\bibitem[{{Denissenkov} \& {Merryfield}(2011)}]{Denissenkov:2011}
{Denissenkov}, P.~A., \& {Merryfield}, W.~J. 2011, \apjl, 727, L8

\bibitem[{{Denissenkov} {et~al.}(2010){Denissenkov}, {Pinsonneault},
  {Terndrup}, \& {Newsham}}]{Denissenkov:2010}
{Denissenkov}, P.~A., {Pinsonneault}, M., {Terndrup}, D.~M., \& {Newsham}, G.
  2010, \apj, 716, 1269

\bibitem[{{Dupret} {et~al.}(2009){Dupret}, {Belkacem}, {Samadi}, {Montalban},
  {Moreira}, {Miglio}, {Godart}, {Ventura}, {Ludwig}, {Grigahc{\`e}ne},
  {Goupil}, {Noels}, \& {Caffau}}]{Dupret:2009}
{Dupret}, M.-A., {Belkacem}, K., {Samadi}, R., {et~al.} 2009, \aap, 506, 57

\bibitem[{{Eggenberger} {et~al.}(2005){Eggenberger}, {Maeder}, \&
  {Meynet}}]{Eggenberger:2005}
{Eggenberger}, P., {Maeder}, A., \& {Meynet}, G. 2005, \aap, 440, L9

\bibitem[{{Eggenberger} {et~al.}(2012){Eggenberger}, {Montalb{\'a}n}, \&
  {Miglio}}]{Eggenberger:2012}
{Eggenberger}, P., {Montalb{\'a}n}, J., \& {Miglio}, A. 2012, \aap, 544, L4

\bibitem[{{Ferrario} \& {Wickramasinghe}(2005)}]{Ferrario:2005}
{Ferrario}, L., \& {Wickramasinghe}, D.~T. 2005, \mnras, 356, 615

\bibitem[{{Gough} \& {McIntyre}(1998)}]{Gough:1998}
{Gough}, D.~O., \& {McIntyre}, M.~E. 1998, \nat, 394, 755

\bibitem[{{Gratton} {et~al.}(2000){Gratton}, {Sneden}, {Carretta}, \&
  {Bragaglia}}]{Gratton:2000}
{Gratton}, R.~G., {Sneden}, C., {Carretta}, E., \& {Bragaglia}, A. 2000, \aap,
  354, 169

\bibitem[{{Greiss} {et~al.}(2014){Greiss}, {G{\"a}nsicke}, {Hermes}, {Steeghs},
  {Koester}, {Ramsay}, {Barclay}, \& {Townsley}}]{Greiss:2014}
{Greiss}, S., {G{\"a}nsicke}, B.~T., {Hermes}, J.~J., {et~al.} 2014, \mnras,
  438, 3086

\bibitem[{{Grosjean} {et~al.}(2014){Grosjean}, {Dupret}, {Belkacem},
  {Montalb{\'a}n}, \& {Samadi}}]{Grosjean:2013}
{Grosjean}, M., {Dupret}, M.-A., {Belkacem}, K., {Montalb{\'a}n}, J., \&
  {Samadi}, R. 2014, in IAU Symposium, Vol. 301, IAU Symposium, ed. J.~A.
  {Guzik}, W.~J. {Chaplin}, G.~{Handler}, \& A.~{Pigulski}, 341--344

\bibitem[{{Heger} {et~al.}(2000){Heger}, {Langer}, \& {Woosley}}]{Heger:2000}
{Heger}, A., {Langer}, N., \& {Woosley}, S.~E. 2000, \apj, 528, 368

\bibitem[{{Heger} {et~al.}(2005){Heger}, {Woosley}, \& {Spruit}}]{Heger:2005}
{Heger}, A., {Woosley}, S.~E., \& {Spruit}, H.~C. 2005, \apj, 626, 350

\bibitem[{{Henyey} {et~al.}(1965){Henyey}, {Vardya}, \&
  {Bodenheimer}}]{Henyey:1965}
{Henyey}, L., {Vardya}, M.~S., \& {Bodenheimer}, P. 1965, \apj, 142, 841

\bibitem[{{Iglesias} \& {Rogers}(1993)}]{Iglesias:1993}
{Iglesias}, C.~A., \& {Rogers}, F.~J. 1993, \apj, 412, 752

\bibitem[{{Iglesias} \& {Rogers}(1996)}]{Iglesias:1996}
---. 1996, \apj, 464, 943

\bibitem[{{Kawaler} \& {Hostler}(2005)}]{Kawaler:2005}
{Kawaler}, S.~D., \& {Hostler}, S.~R. 2005, \apj, 621, 432

\bibitem[{{Koester} {et~al.}(1998){Koester}, {Dreizler}, {Weidemann}, \&
  {Allard}}]{Koester:1998}
{Koester}, D., {Dreizler}, S., {Weidemann}, V., \& {Allard}, N.~F. 1998, \aap,
  338, 612

\bibitem[{{Langer} {et~al.}(1985){Langer}, {El Eid}, \& {Fricke}}]{Langer:1985}
{Langer}, N., {El Eid}, M.~F., \& {Fricke}, K.~J. 1985, \aap, 145, 179

\bibitem[{{Langer} {et~al.}(1983){Langer}, {Fricke}, \&
  {Sugimoto}}]{Langer:1983}
{Langer}, N., {Fricke}, K.~J., \& {Sugimoto}, D. 1983, \aap, 126, 207

\bibitem[{{Lecoanet} \& {Quataert}(2013)}]{Lecoanet:2013}
{Lecoanet}, D., \& {Quataert}, E. 2013, \mnras, 430, 2363

\bibitem[{{Maeder} \& {Meynet}(2000)}]{Maeder:2000}
{Maeder}, A., \& {Meynet}, G. 2000, \araa, 38, 143

\bibitem[{{Marques} {et~al.}(2013){Marques}, {Goupil}, {Lebreton}, {Talon},
  {Palacios}, {Belkacem}, {Ouazzani}, {Mosser}, {Moya}, {Morel}, {Pichon},
  {Mathis}, {Zahn}, {Turck-Chi{\`e}ze}, \& {Nghiem}}]{Marques:2013}
{Marques}, J.~P., {Goupil}, M.~J., {Lebreton}, Y., {et~al.} 2013, \aap, 549,
  A74

\bibitem[{{Mosser} {et~al.}(2012{\natexlab{a}}){Mosser}, {Goupil}, {Belkacem},
  {Michel}, {Stello}, {Marques}, {Elsworth}, {Barban}, {Beck}, {Bedding}, {De
  Ridder}, {Garc{\'{\i}}a}, {Hekker}, {Kallinger}, {Samadi}, {Stumpe},
  {Barclay}, \& {Burke}}]{Mosser:2012a}
{Mosser}, B., {Goupil}, M.~J., {Belkacem}, K., {et~al.} 2012{\natexlab{a}},
  \aap, 540, A143

\bibitem[{{Mosser} {et~al.}(2012{\natexlab{b}}){Mosser}, {Goupil}, {Belkacem},
  {Marques}, {Beck}, {Bloemen}, {De Ridder}, {Barban}, {Deheuvels}, {Elsworth},
  {Hekker}, {Kallinger}, {Ouazzani}, {Pinsonneault}, {Samadi}, {Stello},
  {Garc{\'{\i}}a}, {Klaus}, {Li}, {Mathur}, \& {Morris}}]{Mosser:2012}
---. 2012{\natexlab{b}}, \aap, 548, A10

\bibitem[{{Nielsen} {et~al.}(2013){Nielsen}, {Gizon}, {Schunker}, \&
  {Karoff}}]{Nielsen:2013}
{Nielsen}, M.~B., {Gizon}, L., {Schunker}, H., \& {Karoff}, C. 2013, \aap, 557,
  L10

\bibitem[{{Nordhaus} {et~al.}(2008){Nordhaus}, {Busso}, {Wasserburg},
  {Blackman}, \& {Palmerini}}]{Nordhaus:2008}
{Nordhaus}, J., {Busso}, M., {Wasserburg}, G.~J., {Blackman}, E.~G., \&
  {Palmerini}, S. 2008, \apjl, 684, L29

\bibitem[{{Ouazzani} {et~al.}(2013){Ouazzani}, {Goupil}, {Dupret}, \&
  {Marques}}]{Ouazzani:2013}
{Ouazzani}, R.-M., {Goupil}, M.~J., {Dupret}, M.-A., \& {Marques}, J.~P. 2013,
  \aap, 554, A80

\bibitem[{{Palacios} {et~al.}(2006){Palacios}, {Charbonnel}, {Talon}, \&
  {Siess}}]{Palacios:2006}
{Palacios}, A., {Charbonnel}, C., {Talon}, S., \& {Siess}, L. 2006, \aap, 453,
  261

\bibitem[{{Paxton} {et~al.}(2011){Paxton}, {Bildsten}, {Dotter}, {Herwig},
  {Lesaffre}, \& {Timmes}}]{Paxton:2011}
{Paxton}, B., {Bildsten}, L., {Dotter}, A., {et~al.} 2011, \apjs, 192, 3

\bibitem[{{Paxton} {et~al.}(2013){Paxton}, {Cantiello}, {Arras}, {Bildsten},
  {Brown}, {Dotter}, {Mankovich}, {Montgomery}, {Stello}, {Timmes}, \&
  {Townsend}}]{Paxton:2013}
{Paxton}, B., {Cantiello}, M., {Arras}, P., {et~al.} 2013, \apjs, 208, 4

\bibitem[{{Petrovic} {et~al.}(2005){Petrovic}, {Langer}, {Yoon}, \&
  {Heger}}]{Petrovic:2005}
{Petrovic}, J., {Langer}, N., {Yoon}, S.-C., \& {Heger}, A. 2005, \aap, 435,
  247

\bibitem[{{Rogers} {et~al.}(2013){Rogers}, {Lin}, {McElwaine}, \&
  {Lau}}]{Rogers:2013}
{Rogers}, T.~M., {Lin}, D.~N.~C., {McElwaine}, J.~N., \& {Lau}, H.~H.~B. 2013,
  \apj, 772, 21

\bibitem[{{Shiode} {et~al.}(2013){Shiode}, {Quataert}, {Cantiello}, \&
  {Bildsten}}]{Shiode:2013}
{Shiode}, J.~H., {Quataert}, E., {Cantiello}, M., \& {Bildsten}, L. 2013,
  \mnras, 430, 1736

\bibitem[{{Spada} {et~al.}(2010){Spada}, {Lanzafame}, \& {Lanza}}]{Spada:2010}
{Spada}, F., {Lanzafame}, A.~C., \& {Lanza}, A.~F. 2010, \mnras, 404, 641

\bibitem[{{Spruit}(1999)}]{Spruit:1999}
{Spruit}, H.~C. 1999, \aap, 349, 189

\bibitem[{{Spruit}(2002)}]{Spruit:2002}
---. 2002, \aap, 381, 923

\bibitem[{Suijs {et~al.}(2008)Suijs, Langer, Poelarends, Yoon, Heger, \&
  Herwig}]{larends_Yoon_Heger_Herwig_2008}
Suijs, M. P.~L., Langer, N., Poelarends, A.-J., {et~al.} 2008, \aap, 481, L87

\bibitem[{{Tayar} \& {Pinsonneault}(2013)}]{Tayar:2013}
{Tayar}, J., \& {Pinsonneault}, M.~H. 2013, \apjl, 775, L1

\bibitem[{{Traxler} {et~al.}(2011){Traxler}, {Garaud}, \&
  {Stellmach}}]{Traxler:2011}
{Traxler}, A., {Garaud}, P., \& {Stellmach}, S. 2011, \apjl, 728, L29

\bibitem[{{Zahn} {et~al.}(2007){Zahn}, {Brun}, \& {Mathis}}]{Zahn:2007}
{Zahn}, J.-P., {Brun}, A.~S., \& {Mathis}, S. 2007, \aap, 474, 145

\bibitem[{{Zahn} {et~al.}(1997){Zahn}, {Talon}, \& {Matias}}]{Zahn:1997}
{Zahn}, J.-P., {Talon}, S., \& {Matias}, J. 1997, \aap, 322, 320

\end{thebibliography}
\end{document}